\documentclass[useAMS,usenatbib,usegraphicx]{mn2e}

\usepackage{graphicx,times,epsfig,rotating}

\def\mnras{MNRAS}
\def\aj{AJ}
\def\aap{A\&A}
\def\apj{ApJ}
\def\apjl{ApJ}
\def\apjs{ApJS}
\def\araa{ARA\&A}

\def\nat{Nature}

\newcommand{\kmsend}{\mbox{km s$^{-1}$}}

\newcommand{\msun}{\mbox{M$_{\sun}$ }}
\newcommand{\msunend}{\mbox{M$_{\sun}$}}
\newcommand{\lsun}{\mbox{L$_{\sun}$ }}

\newcommand{\msunyr}{\mbox{M$_{\sun}$yr$^{-1}$ }}
\newcommand{\msunyrend}{\mbox{M$_{\sun}$yr$^{-1}$}}
\newcommand{\htwo}{\mbox{H$_2$}}

\newcommand{\z}{\mbox{$z$}}

\newcommand{\zsim}{\mbox{$z\sim$ }}

\newcommand{\lbol}{L$_{\rm {bol}}$}

\newcommand{\bzk}{\mbox{{\it BzK}}}

\newcommand{\xco}{\mbox{$X_{\rm CO}$}}
\newcommand{\alphaco}{\mbox{$\alpha_{\rm CO}$}}

\newcommand{\fgas}{\mbox{$f_{\rm gas}$}}
\title[On the Gas Fractions of High-Redshift Galaxies]{Galaxy Gas
  Fractions at High-Redshift: The Tension between Observations and
  Cosmological Simulations}

\author[Narayanan, Bothwell \& Dav\'e]{Desika Narayanan$^{1}$\thanks{Bart J. Bok Fellow}\thanks{E-mail:
    dnarayanan@as.arizona.edu}, Matt Bothwell$^{1}$, Romeel Dav\'e$^1$\\$^{1}$Steward
  Observatory, University of Arizona, 933 N Cherry Ave, Tucson, Az,
  85721\\}

\begin{document}

\date{MNRAS Accepted}

\pagerange{\pageref{firstpage}--\pageref{lastpage}} \pubyear{2012}

\maketitle

\label{firstpage}

\begin{abstract}

CO measurements of \zsim 1-4 galaxies have found that their baryonic
gas fractions are significantly higher than galaxies at \z=0, with
values ranging from 20-80\%.  Here, we suggest that the gas fractions
inferred from observations of star-forming galaxies at high-\z \ are
overestimated, owing to the adoption of locally-calibrated CO-\htwo
\ conversion factors (\alphaco).  Evidence from both observations and
numerical models suggest that \alphaco \ varies smoothly with the
physical properties of galaxies, and that \alphaco \ can be
parameterised simply as a function of both gas phase metallicity and
observed CO surface brightness. When applying this functional form, we
find $\fgas \approx 10-40\%$ in galaxies with $M_*=10^{10}-10^{12} \ 
\msunend$.  Moreover, the scatter in the observed \fgas-$M_*$ relation is
lowered by a factor of two.  The lower inferred gas fractions arise
physically because the interstellar media of high-\z \ galaxies have
higher velocity dispersions and gas temperatures than their local
counterparts, which results in an \alphaco \ that is lower than the
\z=0 value for both quiescent discs and starbursts.  We further
compare these gas fractions to those predicted by cosmological galaxy
formation models.  We show that while the canonically inferred gas
fractions from observations are a factor of 2-3 larger at a given stellar mass than
predicted by models, our rederived \alphaco \ values for \z=1-4
galaxies results in revised gas fractions that agree significantly
better with the simulations.

\end{abstract}

\section{Introduction}
\label{section:introduction}

Recent technological advances in (sub)millimetre-wave telescope
facilities have allowed for the detection of star-forming \htwo \ gas
in large numbers of galaxies at high-redshift via the proxy molecule
$^{12}$CO  \citep[hereafter, CO;][see \citet{sol05} for a summary of
  pre-2005
  references]{gre05,tac06,cop08,tac08,bot09,dan09,wag09b,car10,dad10a,dad10b,gen10,rie10a,rie10b,tac10,gea11,cas11,rie11a,rie11b,wan11}.

A major finding from these studies is that, at a given stellar mass,
early Universe galaxies tend to be significantly more gas rich than
their present-day counterparts, with baryonic gas
fractions\footnote{This includes a 36\% correction for Helium.}
(hereafter defined as \fgas = $M_{\rm H2}/(M_{\rm H2}+M_*)$) ranging
from $\sim 20-80\%$ \citep[e.g. ][]{dad10a,tac10}.  This is consistent
with both observational and theoretical results that suggest that
even gas rich disc galaxies at \zsim 2 are able to form stars rapidly
enough that they are comparable to the most extreme merger-driven
starburst events in the local Universe \citep{dad05,hop10,dav10}.

However, there is a tension between the inferred gas fractions of
high-\z \ galaxies and galaxy formation models.  Hydrodynamic
cosmological simulations typically account for the simultaneous growth
of galaxies via accretion of gas from the intergalactic medium
\citep[e.g. ][]{ker03} as well as the consumption of gas by star
formation.  A generic feature of these simulations is that the star
formation rate of ``main sequence galaxies'' (galaxies not undergoing
a starburst event) is roughly proportional to the accretion rate, and
that galaxies tend to have weakly declining gas fractions as their
stellar masses increase.  Broadly, at a given stellar mass, galaxies
in simulations have baryonic gas fractions a factor of 2-3 less than
observed gas fractions.  This is seen both in hydrodynamic
simulations, as well as semi-analytic models \citep{lag11}.  As an
example, \citet{dav10} find very few galaxies in a simulated $\sim
150$ Mpc (comoving) volume with stellar mass $M_* \sim 10^{11}$ \msun
with gas fractions greater than 30\%.  This is in contrast to
observations which infer gas fractions in comparable mass galaxies up
to 80\%.

 %Another way of
%saying this is that modeled massive galaxies with high gas fractions
%tend to consume their gas rapidly \citep[][]{hay11}, and accretion
%rates from the IGM are not rapid enough to offset the star formation
%rate.  Beyond this, simulations predict that \fgas \ should show a
%decreasing trend with increasing stellar mass while CO-based
%observations show a tentative trend at best.

One potential solution is that the inferred gas masses from high-\z
\ galaxies are systematically too large.  \htwo \ masses are
typically calculated using the luminosity of the CO (J=1-0) emission
line\footnote{In fact only a few studies directly observe CO (J=1-0)
  at high-\z.  Typically, higher rotational states are observed, and
  then down-converted to the ground state via an assumption about the
  CO excitation.}, and then converted to an \htwo \ mass via a
CO-\htwo \ conversion factor \alphaco\footnote{\alphaco \ is
  alternatively monikered \xco, or the $X$-factor.  The two are
  related via \xco \ (cm$^{-2} (\rm K-\kmsend)^{-1}$)= $6.3 \times
  10^{19} \alpha_{\rm CO}$ (\msun ${\rm pc}^{-2}$(K-\kmsend)$^{-1}$).
  In this paper, we utilise \alphaco \ as notation for the CO-\htwo
  \ conversion factor.}.  In the literature, \alphaco \ is typically
used bimodally with one value for ``quiescent/disc mode'' star
formation, and a lower value for ``starburst/merger mode''.  In the
Galaxy and Local Group, \alphaco \ is observed to be relatively
constant with an average \alphaco $\approx 6$ \citep{bli07,fuk10}.  In
contrast, dynamical mass modeling of local galaxy mergers suggests
that \alphaco \ should be lower in these galaxies by a factor of 2-10
\citep{dow98,nar11d}.  Despite an observed dispersion in inferred \alphaco
\ values from local mergers, a value of \alphaco$\approx 0.8$ is
typically uniformly applied to these starbursts.

At higher redshifts, it is more unclear which of the two bimodal
values of \alphaco \ to use. For example, for a star-forming disc
galaxy that may be undergoing rapid collapse in $\sim$ kiloparsec
scale clumps and forming stars at rates $> 100$ \msunyr
(i.e. comparable to local galaxy mergers), is the appropriate \alphaco
\ the locally-calibrated ``quiescent/disc'' value, or the
``starburst/merger'' value?  Similarly, should high-redshift
Submillimetre galaxies (SMGs), which are potentially forming stars up
to an order of magnitude faster than local mergers, utilise the
locally-calibrated ``starburst/merger'' value?  Typically,
observational studies use the locally-calibrated quiescent/disc value
for high-\z \ discs, and the local starburst/merger value for high-\z
\ SMGs.  

Recent observational evidence by \citet{tac08,bol08,ler11,gen12,sch12}
and \citet{pap12}, as well as theoretical work by \citet{ost11,
  nar11b, she11a, she11b} and \citet{fel12a} have suggested that
perhaps the picture of a bimodal \alphaco \ is too simplistic, and
that \alphaco \ may depend on the physical environment of the
interstellar medium (ISM).  This picture was expanded upon by
\citet{nar12a} who developed a functional form for the dependence of
\alphaco \ on the CO surface brightness and gas-phase metallicity of a
galaxy.  When applying this model to observations of high-\z
\ galaxies, \citet{nar12a} found that on average, high-\z \ disc
galaxies have \alphaco \ values a factor of a few lower than
present-epoch discs, and high-\z \ SMGs have \alphaco \ values lower
than present-day ultraluminous infrared galaxies (ULIRGs), with
some dispersion.  Physically, this means that for a given
observed CO luminosity, the inferred \htwo \ gas mass should be
systematically less than what one would derive using \alphaco \ values
calibrated to local galaxies.  This owes to warmer and higher velocity
dispersion molecular gas in high-\z \ galaxies which gives rise to
more CO intensity at a given \htwo \ column density.  The model form
for \alphaco \ presented by \citet{nar12a} finds success in matching
local observations of discs and ULIRGs \citep{nar11b}, as well as
observed CO-\htwo \ conversion factors for low metallicity systems.

Building on these results, in this paper, we reexamine CO detections
from high-\z \ galaxies utilising the physically motivated functional
form for \alphaco \ presented in \citet{nar12a}, rather than the
traditional bimodal form.  We compare our results to those of
cosmological hydrodynamic simulations, and show that while the
inferred gas fractions derived from the traditional \alphaco
\ conversion factor are much larger than those predicted by models,
the \citet{nar12a} model form for \alphaco \ brings these values down,
and in reasonable agreement with simulations.  A principle result of
the work we will present is that the typical gas fraction of a high-\z
\ galaxy is typically $\sim 10-40\%$, rather than $\sim40-80\%$ as is
inferred when utilising traditional \alphaco \ values.  
In
\S~\ref{section:observations}, we describe the literature data
utilised here; in \S~\ref{section:results}, we present our main
results, and in \S~\ref{section:summary}, we summarise.

\begin{figure}
\hspace{-1cm}
\includegraphics[angle=90,scale=0.4]{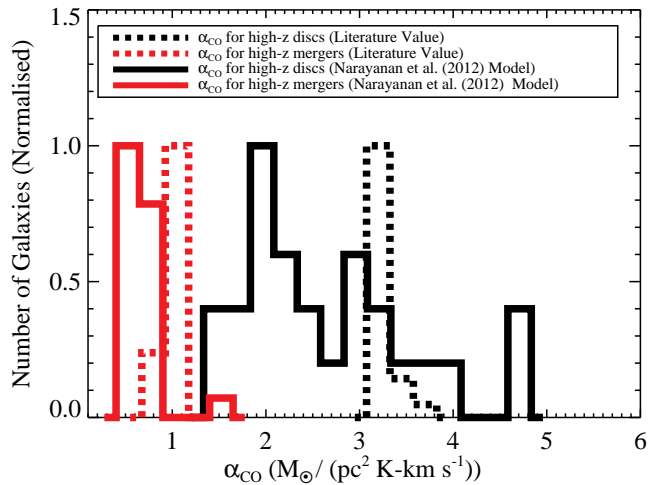}
\caption{Histograms of literature values for \alphaco \ for high-\z
  \ galaxies, as well as those re-derived via
  Equation~\ref{equation:alphaco}.  The dashed lines denote the
  literature values, and solid lines the theoretical \alphaco
  \ values.  The black lines are for high-\z \ discs, and red for
  inferred high-\z \ mergers (SMGs).  On average, the theoretical
  \alphaco \ values are lower than the locally-calibrated (traditional
  literature values) for discs and mergers.  This owes to higher
  velocity dispersion and warmer gas in high-\z \ galaxies which
  drives more CO emission per unit \htwo \ gas mass than in local
  galaxies.\label{figure:alphaco}}
\end{figure}

\begin{figure*}
%\hspace{-1cm}
\includegraphics[angle=90,scale=0.65]{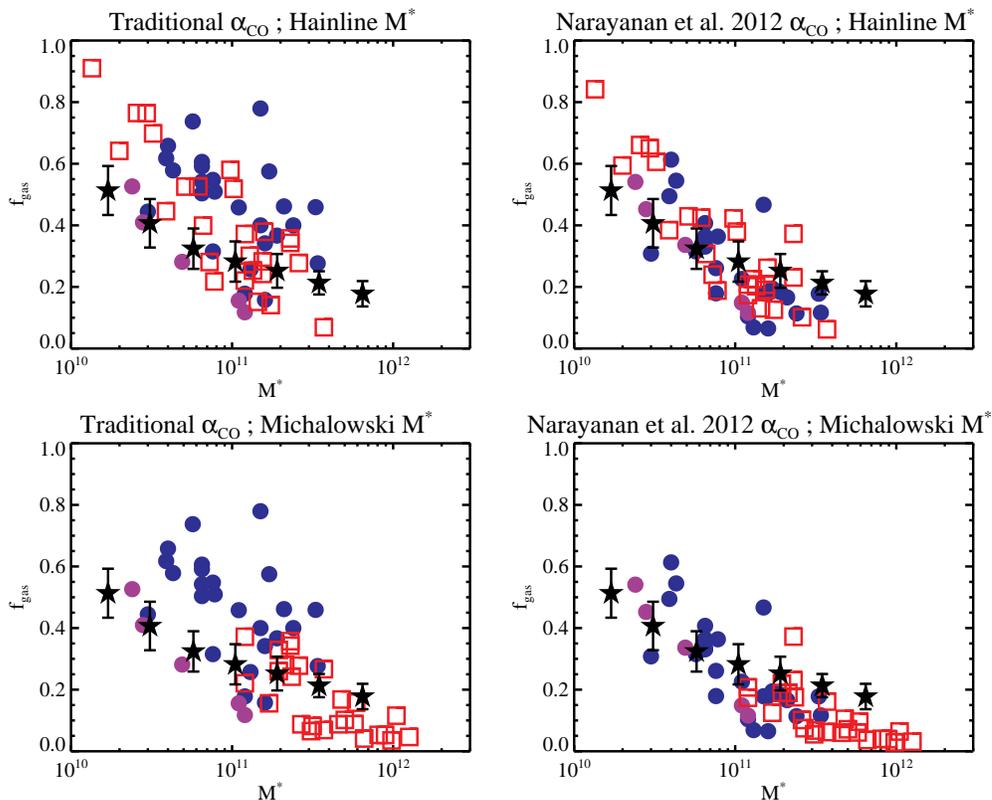}
\caption{Comparison of \fgas \ against stellar mass ($M_{*}$) for
  discs and mergers at high-\z.  The blue filled circles are observed
  high-\z \ discs; the red open squares are observed SMGs (assumed to
  be mergers), and the purple filled circles are optically faint radio
  galaxies (which are of unknown physical form). The black stars show
  results from main-sequence galaxies from the cosmological
  hydrodynamic simulations of \citet{dav10}, with dispersion noted by
  the error bars. The top panels show the results when utilising the
  \citet{hai11} stellar masses, and the bottom panels when utilising
  the \citet{mic09} masses (which are larger by a factor of a few).
  The left panels show the \fgas-$M_*$ relation when utilising the
  traditional locally-calibrated literature \alphaco \ values, and
  right panel shows the effect of our model \alphaco
  \ (Equation~\ref{equation:alphaco}).  When using the
  locally-calibrated \alphaco \ values, the observed gas fractions are
  a factor of 2-3 larger than theoretical gas fractions at a given
  stellar mass.  Because the theoretical \alphaco \ values are
  typically lower than the locally-calibrated values
  (Figure~\ref{figure:alphaco}), the inferred gas mass for a given CO
  luminosity decreases, and brings the gas fractions into better
  agreement with galaxy formation
  simulations.  \label{figure:fgas_mstar}}
\end{figure*}

\section{Methodology}
\label{section:observations}
\subsection{Literature Data}
\label{section:literaturedata}
We examine CO detections of both inferred high-\z \ disc galaxies as
well as Submillimetre Galaxies with
masses ranging from $\sim 10^{10}-10^{12}$ \msun in stellar
mass\footnote{The limits on stellar masses is highly dependent on which
  literature stellar masses for SMGs we use \citep{mic09,hai11}.}.  A
large number of literature SMGs are compiled by \citet{bot12}, and include 16 new detections presented in that
paper.  The compilation by Bothwell et al. includes detections from
\citet{ner03,gre05,tac06,cas09,bot09} and \citet{eng10}.  Other SMGs
included in our work are compiled in \citet{gen10}. The inferred disc
galaxies are taken primarily from the compilation of \citet{gen10} and
\citet{dad10a}.  These include galaxies from the SINS sample
\citep{for09}, as well as \bzk-selected galaxies \citet{dad04}.
Finally, we include optically-faint radio galaxies (OFRGs) with CO detections
from the \citet{cas11} sample. 

In our sample, a large number of the \bzk \ and SINS galaxies have
been imaged and found to have rotationally dominated gas, consistent
with a disc-like morphology.  The SMGs are oftentimes assumed to be
mergers, though there is some debate over this
\citep{dav10,nar09,nar10a,hay10,hay11,hay12a}. The OFRGs are of unknown
origin. As we will discuss in \S~\ref{section:results}, the global
morphology is irrelevant for our model form for \alphaco, and the
general results in this paper.

The observational papers that we draw from had to make a number of
assumptions.  Our philosophy is to simply utilise those assumptions in
this paper, and not make any adjustments to assumed numbers.  The
reason for this is to isolate the effects of applying our model
\alphaco \ on the inferred gas fractions.  For example, as we will
discuss, CO surface brightnesses are required in order to employ the
\citet{nar12a} model for \alphaco.  When direct measurements are
reported, we utilise those. Otherwise, we make the same size
assumption that is made in the paper we draw from.  Similarly, a
number of the detections presented in the aforementioned papers
utilised millimetre-wave telescopes, meaning that the observed
transition is of higher-lying CO lines in the rest frame. Conversion
to the ground state CO (J=1-0) line then occurs via an assumption of
CO excitation.  Again, we simply utilise the conversion from excited
CO lines to CO (J=1-0) as presented in the paper we pull the data
from.  This said, the assumed excitation ladders in the literature are
all relatively similar.

It is worth a quick word on the stellar masses of the SMGs in our
sample.  There is an ongoing literature debate regarding the stellar
masses of high-redshift SMGs.  Specifically, for the {\it same} SMGs,
\citet{mic09} and \citet{hai11} find differing stellar masses by up to
an order of magnitude (with the Hainline masses being lower).  Some
attempts to understand the origin of the discrepancy have been
reported by \citet{mic12}.  In this work, we remain agnostic as to
which stellar masses are ``correct'', and present our results in terms
of both sets of observations when relevant.

%Arguments against the extreme stellar masses from both studies exist.
%It is conceivable that the largest stellar masses ($\sim 10^{12}
%\msunend$) are too large.  Space densities for galaxies of this mass
%are near $\sim10^{-7}$ Mpc$^{-3}$ \citep{mar09}; this is nearly two
%orders of magnitude lower than the observed space density of \zsim 2
%SMGs \citep{tac08,wal08}.  It is unlikely that SMGs are a factor of
%100 more abundant than typical $K$-selected galaxies at the same
%stellar mass.  Similarly, the lowest-mass galaxies ($\sim 10^{10}
%\msunend$) are roughly an order of magnitude lower than is required by
%theoretical SMG formation simulations
%\citep{bau05,nar09,nar10a,nar10b,gon11}.  While this is not firm
%evidence against the lowest mass galaxies, it is compelling that a
%large range of simulation methods (including idealised galaxy mergers,
%semi-analytic models and cosmological hydrodynamic simulations) all
%agree regarding the typical masses of SMGs.

\subsection{Revised CO-\htwo \ Conversion Factors for Observed Galaxies}
\label{section:alphaco}
As discussed in \S~\ref{section:introduction}, we utilise the
functional form of \alphaco \ derived in \citet{nar12a} to
re-calculate the \htwo \ gas masses from the high-\z \ galaxies in our
sample.  In this model, the CO-\htwo \ conversion factor can be
expressed as
\begin{equation}
\label{equation:alphaco}
\alphaco = \frac{10.7 \times \langle W_{\rm CO}\rangle^{-0.32}}{Z'^{0.65}}
\end{equation}
where \alphaco \ has units of \msun pc$^{-2}$/K-\kmsend, $Z'$ is the
gas-phase metallicity in units of solar, and $\langle W_{\rm CO}
\rangle$ is the luminosity-weighted CO intensity over all GMCs in a
galaxy.  While $\langle W_{\rm CO} \rangle$ is a difficult quantity to
observe, in the limit of uniform distribution of luminosity from the
ISM in a galaxy, this reduces to the $L'_{\rm CO}/A$ where $A$ is the
area observed ($L'_{\rm CO}/A$ is the CO surface brightness).  If the
light distribution is actually rather concentrated (and most of the
area observed is in dim pixels), the true surface brightness of the
pixels which emit most of the light will increase, and the true
\alphaco \ will be even lower than what is calculated by
Equation~\ref{equation:alphaco}.  This will cause the gas fractions to
decrease even further from what utilising the \citet{nar12a} model for
\alphaco \ derives, thus enhancing our results.  The CO surface
brightness ($\langle W_{\rm {CO} \rangle}$) serves as a physical
parameterisation for the \htwo \ gas temperature and velocity
dispersion, both of which affect the velocity-integrated CO line
intensity\footnote{For optically thick gas.}  at a given \htwo \ gas
mass.

The functional form for \alphaco \ also depends on a gas-phase
metallicity.  Physically, \alphaco \ varies with the gas-phase
metallicity due to the growth of CO-dark molecular clouds in
low-metallicity gas.  In this regime, the required dust to protect CO
from photodissociating radiation is not present, but the \htwo \ is
abundant enough to self-shield for survival.  For the galaxies in
question, metallicity measurements are typically not available.
Hence, we assume a solar metallicity ($Z'=1$) for all galaxies.  Based
on the \zsim 2 mass-metallicity relation, galaxies of mass $M_* \sim
10^{11}$ \msun typically have metallicities of order solar
\citep{erb06b}.  Thus, an assumption of $Z'=1$ is likely reasonable.
We test the validity of this assumption by examining the effect of
including a stellar mass-metallicity relation.  Following
\citet{erb06b}, we assume all galaxies above $M_* = 10^{11}$ \msun
\ have $Z'=1$, and that the metallicity evolves as ($M_*$)$^{0.3}$ for
lower mass galaxies.  The gas fractions in this test do not deviate by
more than 10\% compared to our assumption that $Z'=1$\footnote{We note
  that there is one galaxy which varies by $\sim 22\%$.  This is the
  lowest mass galaxy in the \citet{hai11} stellar mass
  determinations.}.  This is because of the weak dependence of
metallicity on stellar mass, the weak dependence of \alphaco \ on
metallicity, and the fact that gas fractions depend on stellar mass as
well as gas mass.

It is worth noting that the results in this paper are not entirely
dependent on the model for \alphaco \ given in
Equation~\ref{equation:alphaco}.  A compilation of observations result
in a very similar relation.  \citet{ost11} showed that \alphaco \ and
$\Sigma_{\rm H2}$ are related via a powerlaw in observed galaxies; a
conversion of $\Sigma_{\rm H2}$ to $W_{\rm CO}$ via the relation \xco
= $\Sigma_{\rm H2}/W_{\rm CO}$ (and a linear relationship between \xco
\ and \alphaco) gives an \alphaco-$W_{\rm CO}$ relation that is nearly
identical to the model form in Equation~\ref{equation:alphaco}.
Similarly, while we assume $Z'=1$ for the observed galaxies analysed
in this work, we note that the model power-law relation between
\alphaco \ and $Z'$ is very similar to what has been observed in 
samples of low-metallicity galaxies \citep{bol08,ler11,gen12}.  In this
sense, the forthcoming results in this paper could be derived entirely
from empirical observational evidence.

In Figure~\ref{figure:alphaco}, we plot histograms of the literature
\alphaco \ values used in the observations of the galaxies analysed in
this work (black), as well as our re-derived \alphaco \ values based
on Equation~\ref{equation:alphaco} (red).  We divide the lines into
high-\z \ mergers (where SMGs are assumed to be mergers in this figure
and hereafter) and high-\z \ discs so that the reader can see the
relative difference between our derived \alphaco \ values and the
original ones.  While the range of \alphaco \ values is similar in
both cases, there is significantly more power toward low \alphaco
\ values when utilising our model for both high-\z \ discs and high-\z
\ mergers.

%
%Despite our neutral stance regarding the stellar
%masses, we find it worthwhile to comment that, from a theoretical
%perspective, it is likely that the masses at the extreme ends of the
%Hainline and Micha$\l$owski surveys are too low and too high,
%respectively.  

%The low end of the Hainline masses approach $\sim 10^{10}$ \msunend,
%and are no different on average than the typical \bzk \ disc galaxy
%selected at \zsim 2.  Whether SMGs be primarily fueled by
%main-sequence disc galaxies which are fed by cold streams from the IGM
%\citep{dav10}, major mergers \citep{bau05,nar09,nar10a,nar10b,gon11},
%Or a combination of the two \citep[][Hayward et al. in
%  prep.]{hay10,hay11}, all theoretical models predict that the average
%SMG should have a stellar mass of $\sim 10^{11} \msun$.  While this is
%not quantitative proof that SMGs are unlikely to be at masses this
%low, and it is conceivable that no theoretical model has yet precisely
%reproduced the observed properties of SMGs, it is compelling that the 
%aforementioned models cover a large range of modeling techniques: they
%include idealized merger simulations, semi-analytic models,
%semi-empirical models and full cosmological hydrodynamic simulations.

\begin{figure}
\hspace{-1cm}
\includegraphics[angle=90,scale=0.4]{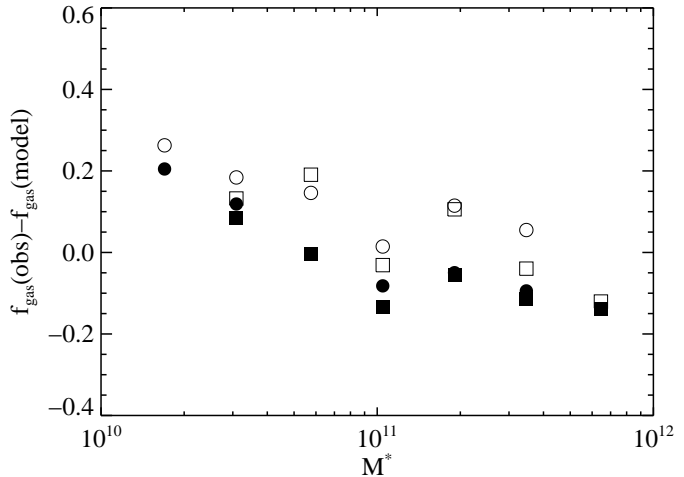}
\caption{Difference between observed gas fractions and cosmological
  simulation gas fractions as a function of stellar mass.  The squares
  are the results when utilising the \citet{mic09} data, and circles
  from the \citet{hai11} data.  The open symbols denote the results
  when the observed gas fractions are calculated utilising traditional
  \alphaco \ values, and the filled symbols correspond to values
  calculated from the \citet{nar12a} form for \alphaco.  Our
  functional form for \alphaco \ results in lower gas fractions and,
  generally, better agreement between the simulations and
  observations. \label{figure:fgas_residuals}}
\end{figure}

\begin{figure*}
\includegraphics[angle=90,scale=0.65]{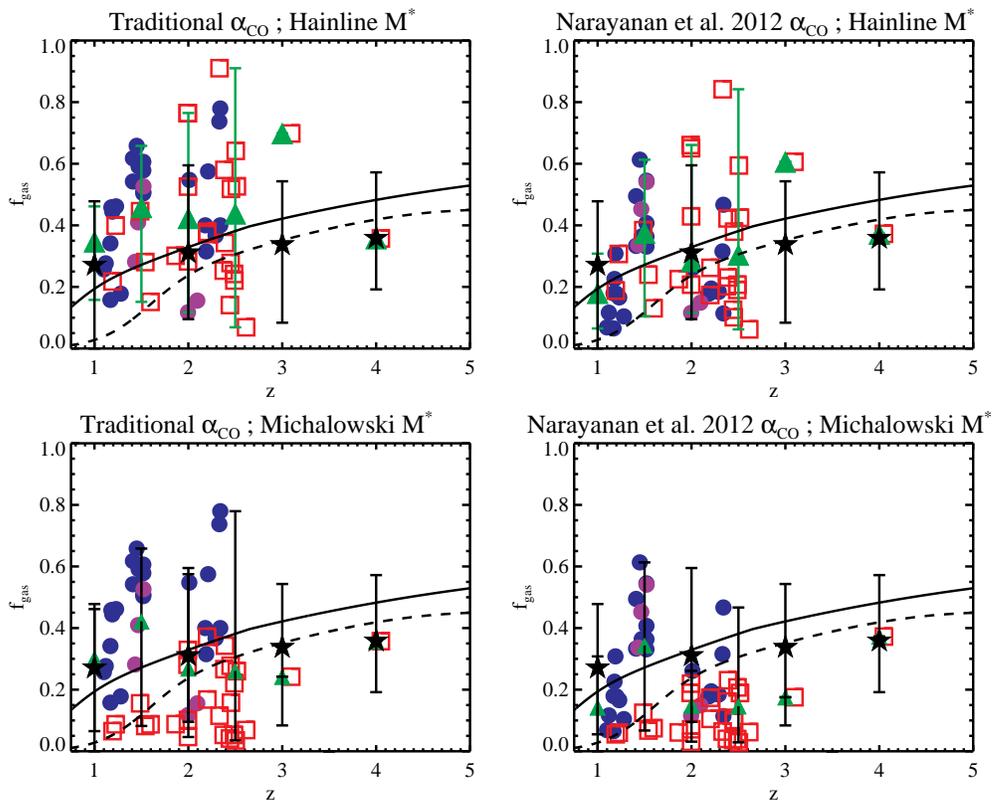}
%\vspace{-2.25in}
\caption{Comparison of \fgas \ against redshift. Filled circles and
  open squares are same as Figure~\ref{figure:fgas_mstar}.  Green
  triangles are median and range of data binned in bins of $\Delta z =
  0.5$, and are presented to highlight the trends of $f_{\rm gas}$ with redshift. Solid
  line and dashed line denotes predicted evolution for galaxies in
  (\z=0) haloes of $M=10^{13}$ and $10^{14}$\msunend, respectively
  \citep{dav11}.  The stars show the mean value and 1$\sigma$ scatter
  for all simulated galaxies with $10.5 < {\rm log (M_*)} < 11.5$.
  Both observations and model predict that the mean gas fractions of
  galaxies should rise with redshift, though utilising our model
  \alphaco \ (right panels) versus the traditional literature values
  (left panels) brings the normalisation of the observed and
  theoretical \fgas-\z \ relation in better
  agreement. \label{figure:fgas_z}}
\end{figure*}

\section{Results and Discussion}
\label{section:results}

We first examine the effect of modifying \alphaco \ from the
traditional bimodal values to the \citet{nar12a} model on the \fgas-$M_*$
relation in high-\z \ galaxies. In Figure~\ref{figure:fgas_mstar}, we
show the \fgas-$M_*$ relation for all galaxies in our sample utilising
both the \citet{mic09} and \citet{hai11} stellar masses.  The left
panels show the relationship for the observed galaxies when using the
traditional bimodal \alphaco, and the right panels when applying
Equation~\ref{equation:alphaco}.  In order to compare with galaxy
formation simulations, we overlay the mean \fgas-$M^*$ relation from
the cosmological hydrodynamic calculations of \citet[][]{dav10}
denoted by stars.  The error bars in the stars denote the range in
possible \fgas \ values for the simulated galaxies in a given stellar
mass bin.  The simulated galaxies mostly represent ``main-sequence'' galaxies
which are not typically undergoing a starburst event.

When examining the left panels in Figure~\ref{figure:fgas_mstar}, it
is evident that the observed galaxies all have substantially higher
gas fractions at a given stellar mass than the simulations.  While
lower-mass systems can be biased to somewhat higher gas fractions
because they are (in part) selected by far-infrared luminosity, we
have tried applying similar cuts to simulations and find that this
cannot explain the difference.  In contrast, when applying a CO-\htwo
\ conversion factor which varies smoothly with physical environment,
the inferred \htwo \ gas masses from the CO line measurements drop and
come into better agreement with the simulations.  Depending on the
stellar mass adopted for the SMGs, the gas fractions can drop by up to
a factor of 3 for a given galaxy.  

The reason for the drop in gas fraction when using the \citet{nar12a}
model for \alphaco \ versus the traditional bimodal value is due to
the typical environments of high-\z \ galaxies.  The gas fractions of
high-\z \ discs are typically large enough that, absent substantial
internal feedback, large $\sim$kpc-scale clumps of gas become unstable
and fragment \citep{spr05a,cev10,hop11a}.  These clumps can have large
internal velocity dispersions ($\sim 50-100 \ \kmsend$), and warm gas
temperatures owing to high star formation rates \citep[$\ga 100
  \ \msunyrend$;][]{nar11a}.  High velocity dispersions and warm gas
causes increased CO line luminosity for a given \htwo \ gas mass, and
reduces \alphaco \ \citep{nar11b}.  Because of this, in our model,
high-\z \ disc galaxies tend to have lower\footnote{Increased UV
  photons produced in high SFR galaxies do have the potential to
  photodissociate CO.  However, these galaxies tend to have large dust
  to gas ratios.  Increased dust columns allow GMCs to reach $A_{\rm
    V} \approx 1$ quickly, and shield CO from photodissociation
  throughout the bulk of the GMC \citep{nar12a}.} \alphaco \ values
than the traditional present-epoch ``quiescent/disc'' value (though
larger than the traditional present-epoch ``starburst/merger'' value;
Figure~\ref{figure:alphaco}).  The mean derived \alphaco \ for high-\z
\ discs is 2.5, approximately half that of the \citet{dad10a} and
\citet{mag11} measurements of high-\z \ \bzk \ galaxies.

A similar effect is true for high-\z \ starburst galaxies.  Owing to
extreme star formation rates \citep[potentially up to a thousand
  \msunyrend;][]{nar12b}, the gas temperatures and velocity
dispersions in violent \zsim 2 mergers exceed those of even
present-day ULIRGs.  Hence, the average \alphaco \ is lower than the
average ULIRG value today.  Our average derived value for the high-\z
\ galaxies in our sample is $\alphaco \approx 0.5$.   \citet{mag11}
finds an upper limit of the \alphaco \ of a \z=4 SMG of 1, and
\citet{tac08} finds a reasonable fit to their observed SMGs with an
\alphaco \ of unity.

The combined effect of our modeling is that \alphaco \ for high-\z
\ discs is typically lower than that of the traditional \z=0
``quiescent/disc'' value, and \alphaco \ for high-\z \ starbursts is
lower than that of the traditional \z=0 ``starburst/merger'' value
(Figure~\ref{figure:alphaco}).  Employing our model \alphaco
\ consequently lowers gas masses, and brings gas fractions into better
agreement with cosmological galaxy formation models.  This is
quantitatively shown in Figure~\ref{figure:fgas_residuals}, where we
plot the residuals between the observed data and models for both the
traditional \alphaco, as well as that derived from the \citet{nar12a}
functional form.

The usage of our model form of \alphaco \ reduces the scatter in the
\fgas-$M_*$ relation in observed galaxies by a factor $\sim 2$ at a
given $M_*$. To calculate the reduction in scatter, we compare the
standard deviation in galaxy gas fractions within a limited range of
stellar masses ($M_* = [5\times10^{10},10^{11}] \msunend$).  Much of
the scatter in the original \fgas-$M_*$ relation arises from using the
bimodal \alphaco \ values.  In contrast, our model form of \alphaco
\ varies smoothly with the physical conditions in the ISM in a galaxy,
and has no knowledge as to whether or not the global morphology of a
galaxy is a merger or a disc. So, if some high-\z \ disc galaxies
actually have physical conditions in their ISM comparable to
starbursts, then their \alphaco \ values will be lower than the
canonical ``quiescent/disc'' \alphaco \ (Figure~\ref{figure:alphaco}).
The vice-versa is true for high-\z \ SMGs and OFRGs.  When accounting
for the continuum in physical properties in the ISM of high-\z
\ galaxies (rather than binning them bimodally), the scatter in the
observed \fgas-$M_*$ relation reduces substantially.  The
correlation coefficient between the observed gas fractions and modeled
ones increases by $\sim 10\%$ (from $\sim 0.9$ to $0.98$ for both the
Michalowski and Hainline masses.

The usage of the \citet{mic09} masses result in observed \fgas \ below
the simulations for the highest mass galaxies.  This could reflect
either an overestimate of masses, or perhaps physical processes that
are neglected in the \citet{dav11} simulations where most SMGs (the
most massive galaxies) are quiescent, main-sequence objects.
Potential neglected physical processes include starbursts which may
deplete gas \citep[e.g. ][]{nar09,nar10b,nar10a,hay11}, or a stage of
gas consumption without replenishment (that ultimately ends in passive
galaxies).

Utilising our model \alphaco \ additionally results in better
agreement between the observed cosmic evolution of the gas fraction of
galaxies with redshift and modeled evolution.  In
Figure~\ref{figure:fgas_z}, we plot the observed gas fractions of the
galaxies in our literature sample against their redshifts. To help
guide the eye, we overplot the mean values (with dispersion) in
redshift bins of 0.5.  We show the predicted values from the analytic
model of \citet{dav11} for halos of mass (at \z=0) $10^{13}$ and
$10^{14}$ \msun by the solid and dashed lines respectively, and the
simulated points from \citet{dav10} by the stars.

When comparing the mean observed values to the predictions from
analytic arguments and cosmological simulations, we again see that
when using the traditional \alphaco \ calibrated to local values, the
inferred gas fractions are significantly higher than the predictions
from models.  When applying the \citet{nar12a} model for \alphaco, the
mean values come into better agreement with simulations. The scatter
(as measured as the standard deviation in gas fractions between
$\z=0-3$) decreases by $\sim 25\%$ when utilising our model form for
\alphaco \ as compared to the traditional values\footnote{We note that
  to properly evaluate the evolution of the gas fraction of galaxies
  with redshift, one would ideally examine the same limited stellar
  mass range at each redshift interval.  Given the limited number of
  CO detections at high-\z, however, this is currently infeasible.  An
  examination of the galaxies in Figure~\ref{figure:fgas_mstar} shows
  that the majority of our galaxies reside in a stellar mass range of
  log$(M_*) = 10.5-11.5$.  The lack of clean sample selection is
  evident in the marginal increase in the already weak correlation
  coefficients: the correlation coefficient increases from $0.08$ to
  $0.14$ for the Hainline masses, and remains roughly constant at
  $\sim 0.15$ for the Michalowski masses.  Forthcoming work will
  address the cosmic evolution of galaxy gas fractions in more
  detail.}.

It is important to note that it is the normal star-forming galaxies
(e.g. the \bzk \ galaxies represented by the filled blue circles) that
come into better agreement with the simulations.  On the other hand,
SMGs, represented by the open red squares in
Figure~\ref{figure:fgas_z}, have lower gas fractions than the models
predict. This is because SMGs are not typical galaxies at high-\z, but
rather rare massive outliers, and therefore not reflected in the
predictions for an average galaxy at high-\z.

%Similarly, when utilising the \citet{hai11} masses, the observed gas
%fractions all appear to be systematically larger than the simulations,
%even when utilising our model \alphaco.  This may arise due to the
%fact that at low masses, there is a selection effect towards higher
%SFR galaxies, and thus higher \fgas.  

Comparing the results of this paper to other galaxy formation models
is nontrivial.  For example, when comparing to galaxies above
\lbol$>10^{11}$ \lsun, the gas fractions returned from our model are
significantly lower than those predicted by recent semi-analytic
models (SAMs).  \citet{lag11} utilised the Durham SAM to predict the
\htwo \ content in galaxies over cosmic time.  As shown by
\citet{bot12}, these models substantially over predict the \htwo \ gas
fraction as a function of redshift.  However, comparing to galaxies
above \lbol$>10^{12}$ \lsun \ produces better agreement, however
(C. Lagos, private communication).  \citet{pop12} utilised an indirect
methodology to derive the \htwo \ content in observed galaxies.  By
inverting the Schmidt relation, and using the \citet{bli06}
pressure-based prescription for deriving the \htwo/HI ratio, these
authours found a gas fraction-stellar mass relation in good agreement
with those derived from CO measurements.  Implicit in this model,
however, is an assumption of an \alphaco \ conversion factor in
setting the normalisation of the observed Schmidt relation. In this
sense, the measurement is not entirely independent of the methods used
in CO-derived gas fractions.

\section{Summary}
\label{section:summary}

Observed baryonic gas fractions from high-redshift galaxies as
inferred from CO measurements are typically higher at a given stellar
mass or redshift than cosmological galaxy formation models would
predict.  These differences can amount to a factor of 2-3 in gas
fraction.  

We suggest that the observed gas fractions are overestimated due to
the usage of locally-calibrated CO-\htwo \ conversion factors
(\alphaco).  If \alphaco \ scales inversely with the CO surface
brightness from a galaxy (as both numerical models and empirical
observational evidence suggest), then both high-\z \ disc galaxies
and high-\z \ mergers will have lower average \alphaco \ values than
their \z=0 analogs.  This means that for a given CO luminosity,
there will be less underlying \htwo \ gas mass.  

Applying a functional form for \alphaco
\ (Equation~\ref{equation:alphaco}) decreases the inferred \htwo \ gas
masses by a factor of $\sim 2-3$, and brings them into agreement with
cosmological galaxy formation models.  Similarly, the usage of our
model \alphaco \ reduces the scatter in the observed \fgas-$M_*$
relation by a comparable amount.  Galaxy gas fractions decrease
monotonically with increasing stellar mass, while the average gas
fraction of galaxies in a given stellar mass range increases with
redshift.

\section*{Acknowledgements} 
DN acknowledges support from the NSF via grant AST-1009452 and thanks
Claudia Lagos and Gerg\"{o} Popping for helpful conversations.  RD was
supported by the NSF under grant numbers AST-0847667 and AST-0907998.
We additionally thank the anonymous referee for helpful suggestions
that improved the presentation of these results.  Computing resources
were obtained through grant number DMS-0619881 from the National
Science Foundation.

\end{document}